\documentclass[aps,prl,amssymb,twocolumn,10pt,floatfix,superscriptaddress]{revtex4-2}
\usepackage[mathscr]{euscript}
\usepackage{bm}
\usepackage{exscale}
\usepackage{textcomp}
\usepackage{comment}
\usepackage{slashed}
\usepackage{tabularx}
\usepackage{wrapfig}
\usepackage{makecell}
\usepackage{amsmath,amsfonts,amsthm,amsbsy,mathtools} 
\usepackage{graphicx} 
\usepackage{bbold} 
\usepackage{physics}
\usepackage{hyperref} 
\usepackage[dvipsnames]{xcolor} 
\usepackage{enumitem} 
\usepackage{pst-all}
\usepackage[caption=false,
    labelformat=simple,
    listofformat=subsimple]{subfig}
\usepackage{makecell}

\definecolor{MathematicaPink}{RGB}{255, 128, 128}
\definecolor{MathematicaCyan}{RGB}{0, 179, 255}
\definecolor{MathematicaPurple}{RGB}{128, 0, 128}

\def\hc{\text{H.c.}}
\newcommand{\pbr}[1]{\left(#1\right)}
\newcommand{\sbr}[1]{\left[#1\right]}
\newcommand{\cbr}[1]{\left\{#1\right\}}

\def\Z{\mathbb{Z}}

\def\O{\mathcal{O}}

\def\S{\mathcal{S}}
\def\U{\mathcal{U}}

\def\da{\dagger}

\def\pa{\partial}

\def\ra{\rightarrow}

\def\al{\alpha}

\def\de{\delta}
\def\De{\Delta}

\def\Ga{\Gamma}
\def\la{\lambda}
\def\ka{\kappa}

\def\pa{\partial}

\def\vp{\varphi}

\def\si{\sigma}

\def\th{\theta}
\def\Th{\Theta}

\hypersetup{
    unicode=false,          
    pdftoolbar=true,        
    pdfmenubar=true,        
    pdffitwindow=true,     
    pdfstartview={FitH},    
    pdftitle={ZM-in-FP},    
    pdfauthor={},     
    pdfsubject={},   
    pdfcreator={},   
    pdfproducer={}, 
    pdfkeywords={zero mode} {non-Abelian anyons} {symmetry fractionalization} {symmetry defects}, 
    pdfnewwindow=true,      
    colorlinks=true,       
    linkcolor=MathematicaPurple, 
    citecolor=blue,        
    filecolor=MathematicaPink,      
    urlcolor=MathematicaCyan      
} 

\begin{document}
\title{Interferometric Signatures of Zero Modes in Fractional Quantum Hall-Superconductor Heterostructures}
\author{Junyi Cao}
\affiliation{Department of Physics, University of Illinois Urbana-Champaign, Urbana, IL 61801, USA}
\affiliation{The Anthony J. Leggett Institute for Condensed Matter Theory, University of Illinois Urbana-Champaign, 1110 West Green Street, Urbana, IL 61801, USA}
\author{Ramanjit Sohal}
\affiliation{Pritzker School of Molecular Engineering, University of Chicago, 5640 S Ellis Ave, Chicago, IL 60637, USA}
\author{Angela Kou}
\affiliation{Department of Physics, University of Illinois Urbana-Champaign, Urbana, IL 61801, USA}
\affiliation{Materials Research Laboratory, University of Illinois Urbana-Champaign, Urbana, IL 61801, USA}
\affiliation{Holonyak Micro and Nanotechnology Lab, University of Illinois Urbana-Champaign, Urbana, IL 61801, USA}
\author{Eduardo Fradkin}
\affiliation{Department of Physics, University of Illinois Urbana-Champaign, Urbana, IL 61801, USA}
\affiliation{The Anthony J. Leggett Institute for Condensed Matter Theory, University of Illinois Urbana-Champaign, 1110 West Green Street, Urbana, IL 61801, USA}
\begin{abstract}
    Fractional quantum Hall-superconductor (FQH-SC) heterostructures are predicted to host defect-bound parafermion zero modes (PZMs). 
    We propose two related configurations to probe their fusion structure. In a Josephson junction coupled to a single quantum point contact (QPC), quasiparticle tunneling switches the defect fusion channel, producing stochastic transitions between branches of the fractional Josephson spectrum. Embedding the junction in a two-QPC Fabry-P\'{e}rot interferometer provides a complementary probe. Weak zero mode tunneling produces fusion-channel-dependent interference, while strong tunneling makes the interferometer probe a superposition of fusion channels and strongly suppresses the signal: in the
    topological limit it vanishes exactly, revealing the defects' non-Abelian nature even when the parent FQH state is Abelian.
\end{abstract}
\date{August 17, 2026}
\maketitle

\textit{Introduction---}Topologically protected degeneracies of certain quantum states underlie many proposals for fault-tolerant quantum computation. In two-dimensional topological phases of matter, non-Abelian anyons provide one route to realizing such degenerate Hilbert spaces, with braiding yielding unitary operations in the associated state space 
\cite{kitaev-2003,nayak-2008}. One potential difficulty of this theoretically appealing proposal is that anyons are dynamical excitations and are thus difficult to control experimentally. An alternative approach is provided by the creation and manipulation of non-Abelian defects that localize \emph{parafermion zero modes} (PZMs) \cite{fendley-2012}. Fractional quantum Hall-superconductor (FQH-SC) heterostructures have been proposed as an experimentally realizable platform for creating such defects \cite{clarke-2013,lindner-2012,cheng-2012}. Because the defects are non-dynamical, the bound PZMs are expected to be more readily addressable in experiments than their anyonic counterparts. 

Recent advances in engineering fractionalized topological phases in two-dimensional van der Waals materials \cite{du-2009,bolotin-2009,spanton-2018,anderson-2023}, and in particular the discovery of the fractional quantum anomalous Hall effect in these systems \cite{cai-2023,park-2023,zeng-2023,lu-2024}, have brought renewed interest in these proposals. Indeed, graphene and transition-metal-dichalcogenide-based systems now allow fractional quantum Hall (FQH) states, superconducting contacts, and gate-defined nanostructures to be combined within closely related experimental settings \cite{lee-2017,gul-2022,jia-2024,han-2025,xu-2025-1}. At the same time, fractional quantum anomalous Hall (FQAH) states provide a path toward FQH physics without an external magnetic field \cite{cai-2023,zeng-2023,park-2023,lu-2024}, thereby circumventing the difficulty of combining strong magnetic fields with superconductivity.

\begin{figure}
    \centering
    \subfloat[]{%
        \includegraphics[width=0.43\columnwidth]{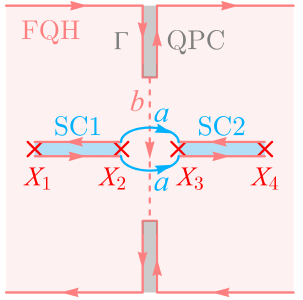}
        \label{fig:qpcj1}
    }
    \subfloat[]{%
        \includegraphics[width=0.57\columnwidth]{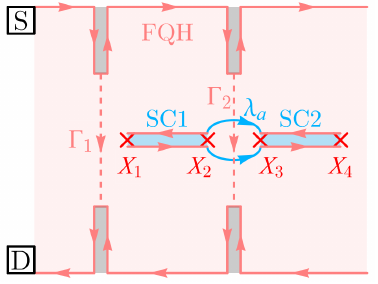}
        \label{fig:FP}
    }
    \caption{Proposed device configurations. (a) An FQH-SC Josephson junction coupled to a single QPC. The superconducting regions SC1 and SC2 (cyan) terminate at defects $X_1,\ldots,X_4$ (red). QPC tunneling transfers an anyon $b$ of the parent FQH state across the constriction with amplitude $\Gamma$. The two $a$ lines represent tunneling of an ``anyon Cooper pair'' between SC1 and SC2; projected onto the defect Hilbert space, this process gives the zero mode tunneling operator $\alpha_a^{(2)\dagger}\alpha_a^{(3)}$ with amplitude $\lambda_a$. (b) An FQH-SC Josephson junction embedded in a two-QPC Fabry-P\'{e}rot interferometer. The QPC tunneling amplitudes are $\Gamma_1$ and $\Gamma_2$, while S and D denote the source and drain.}
    \label{fig:qpcj_FP}
\end{figure}
\begin{table*}[!t]
    \centering
    \begin{tabular}{||c|c|c||}
    \hline 
    Parent FQH state& Anyons &Zero mode structure\\
    \hline
    Fermionic Laughlin state, $\nu=1/m$ ($m$ odd)& $\cbr{j|j\in\Z_m}$ & $\Z_{2m}$\\
    Jain/Halperin state, $\nu=2/5$ & $\cbr{j|j\in\Z_5}$ &$\Z_{5}$\\
    Jain/Halperin state, $\nu=2/3$ & $\cbr{j|j\in\Z_3}$ & $\Z_{3}$\\
    Hole-conjugated Jain/Halperin state, $\nu=3/5$ & $\cbr{j|j\in\Z_5}$ & $\Z_{10}$\\
    Halperin $(3,3,1)$ state, $\nu=1/2$& $\cbr{j|j\in\Z_8}$ &$\Z_{4}$\\
    Fermionic Moore-Read state, $\nu=1/m$ ($m$ even) &$\cbr{a_r|a\in\{1,\chi,\si\},~r\in\Z_{4m}}$ &$\Z_2^{(n)}\times\Z_{2m}$\\
    \hline
    \end{tabular} 
    \caption{Anyon labels and defect zero mode structures for representative FQH-SC heterostructures. For fermionic MR state, physical anyons are restricted by locality with respect to the electron: $1_r$ and $\chi_r$ have even $r$, while $\si_r$ has odd $r$. Abelian parent FQH states support $\Z_N$-type parafermion zero modes, whereas a Moore-Read FQH-SC contains an additional neutral $\Z_2$ zero mode sector arising from non-Abelian fusion channels, denoted by $\{1,\chi\}\simeq\Z_2^{(n)}$. Whether fermion parity enlarges the zero mode structure depends on the number and chirality of the propagating edges \cite{barkeshli-2013,barkeshli-2013-2}.}
    \label{tab:zms}
\end{table*}
Building on extensive theoretical work on defect-bound zero modes and their experimental signatures \cite{kitaev-2001,bombin-2010,clarke-2013,lindner-2012,cheng-2012,barkeshli-2012,you-2012,fendley-2012,you-2013,barkeshli-2013-genon,burrello-2013,barkeshli-2013,barkeshli-2013-2,barkeshli-2014-1,teo-2014-1,teo-2014-2,clarke-2014,barkeshli-2014,teo-2015,alicea-2016,chen-2016,ebisu-2017,barkeshli-2019,groenendijk-2019,schiller-2020,nielsen-2022,teixeira-2022,calzona-2023,nielsen-2023,schiller-2023,cao-2024,wen-2024,bollmann-2026}, we propose two closely related experimental configurations for realizing and probing zero modes in an FQH-SC heterostructure. Both are based on an FQH-SC Josephson junction coupled to QPCs, as shown in Fig.~\ref{fig:qpcj_FP}. The single-QPC configuration in Fig.~\ref{fig:qpcj1} probes fusion-channel switching through the fractional Josephson spectrum, whereas Fig.~\ref{fig:FP} embeds the junction in a two-QPC Fabry-P\'{e}rot interferometer \cite{chamon-1997,fradkin-1998,bonderson-2006,stern-2006,bonderson-2009,bishara-2009,stern-2010,halperin-2011,nakamura-2019,nakamura-2020,feldman-2021,feldman-2022,nakamura-2023,werkmeister-2025,samuelson-2026} that probes the defect fusion channels through interference. The former can be obtained from the latter by suppressing tunneling at one of the two QPCs.

The operating principles of the two configurations are as follows. In the single-QPC configuration, the mutual statistical linking between quasiparticle tunneling across the QPC and zero mode tunneling across the Josephson junction shifts the defect fusion channel, producing stochastic transitions between branches of the fractional Josephson spectrum. In the two-QPC configuration, Fabry-P\'{e}rot interference probes the defect fusion channels through the two-path interference signal. When zero mode tunneling across the junction is negligible, the fusion channel probed by the interferometer is well defined and the device exhibits interference. In the strong zero mode tunneling limit, zero mode tunneling instead fixes the fusion channel across the junction, so the interferometer probes a superposition of fusion channels and the interference is strongly suppressed. In the ideal topological limit, the interference vanishes exactly, providing a signature of the defects' non-Abelian fusion structure even when the parent FQH state is Abelian. Together, the two configurations allow the defect fusion channels to be switched and probed. Below, we first analyze zero mode tunneling across the Josephson junction and QPC-induced fusion-channel switching in Fig.~\ref{fig:qpcj1}, and then turn to the two-QPC interferometric response in Fig.~\ref{fig:FP}.

\textit{Zero mode tunneling across the Josephson junction---}We first consider the Josephson junction between the superconductors labeled SC1 and SC2 in Fig.~\ref{fig:qpcj1}. The superconducting regions proximity-couple and gap the edge modes of the parent FQH state, forming FQH-SC segments whose endpoints define defects $X_1,\ldots ,X_4$, with $X_2$ and $X_3$ forming the Josephson junction. Each defect $X_i$ localizes zero modes whose algebra encodes the nonlocal Hilbert space of multiple defects \cite{clarke-2013,cheng-2012,lindner-2012,barkeshli-2013,barkeshli-2013-genon,barkeshli-2013-2,SM}. For FQH-SCs with an Abelian Laughlin parent state, the resulting zero modes $\alpha$ satisfy a $\Z_N$ parafermionic algebra \cite{fradkin-1980} with the parent-state-dependent values of \(N\) summarized in Table \ref{tab:zms} \footnote{Zero mode structures in Abelian FQH-SCs are discussed in Refs.~\cite{barkeshli-2013,barkeshli-2013-2}. A general derivation of the zero mode structures, including the non-Abelian examples summarized here, is provided in Ref.~\cite{cao-2026}.}. In fermionic FQH-SCs, fermion parity distinguishes additional defect fusion channels and can enlarge the zero mode structure \cite{barkeshli-2013}. For example, although the physical quasiparticles of a fermionic Laughlin state at filling $\nu=1/m$ are labeled by $\Z_m$, its defect fusion channels form $\Z_{2m}$, giving a $\Z_{2m}$ zero mode structure \cite{SM}. 

Coupling zero modes across the junction lifts the degeneracy among the $X_2\times X_3$ fusion channels. At low energies, we describe this coupling by \cite{kitaev-2001,clarke-2013,cheng-2012,lindner-2012}:
\begin{align}
    H_{\text{tunneling}}=\sum_a \la_a \al^{(2)\da}_a\al^{(3)}_a+\hc, \label{eqn:hamiltonian}
\end{align}
where $a$ labels the topological sector transferred by the zero mode tunneling process, and $\la_a$ is the corresponding zero mode tunneling amplitude, assumed to be much smaller than the proximitized superconducting gap. The sum includes all symmetry-allowed zero mode tunneling processes. Physically, this describes the tunneling of ``anyon Cooper pairs" from SC1 to SC2, as shown in Fig.~\ref{fig:qpcj1}. We focus on the tunneling processes labeled by the minimally charged sectors, which are expected to dominate over composite zero mode tunneling processes.

For a $\Z_N$ zero mode structure, the tunneling spectrum is obtained by diagonalizing the tunneling operator $\al^{(2)\da}_1\al^{(3)}_1$ in the $X_2\times X_3$ fusion basis, where we use $\al_1$ to denote the zero mode associated with the elementary nontrivial charge sector. 
We choose the basis to be $\cbr{\ket{l}}$, where $l\in\Z_N$ labels the fusion channel of $X_2\times X_3$, such that \footnote{An additional $e^{i(N-1)\pi/N}$ phase is required in the eigenvalue of the zero mode tunneling operator to maintain the parafermion algebra \cite{fradkin-1980,clarke-2013}. Here we retain only the fusion-channel-dependent part of the eigenvalue for simplicity, as the additional phase shifts the tunneling spectrum for all fusion channels.}
\begin{align}
    \al_1^{(2)\da}\al_1^{(3)}\ket{l}=e^{\frac{2\pi i}{N}l}\ket{l}\,.\label{eqn:zm_tunnel}
\end{align}
Let $\phi_{\rm SC}=\phi_{2}-\phi_{1}$ denote the superconducting phase difference across the junction. Because the zero mode tunneling operator $\alpha_{1}^{(2)\dagger}\alpha_{1}^{(3)}$ transfers charge \(2e/N\) between the two superconductors, gauge invariance requires its tunneling amplitude to acquire the phase $e^{-i\phi_{\text{SC}}/N}$. Thus, $\la_1=\abs{\la_{1}}e^{-i\frac{\phi_{\text{SC}}}{N}}$. Combining this phase dependence with Eq.~\eqref{eqn:zm_tunnel}, the fusion-channel-dependent spectrum thus exhibits a fractional Josephson effect \cite{kitaev-2001,clarke-2013,cheng-2012,lindner-2012}, for which each fixed-$l$ branch is $2\pi N$-periodic in $\phi_{\text{SC}}$:
\begin{align}
    E_l(\phi_{\text{SC}})&=2\abs{\la_1}\cos(\frac{2\pi l}{N}-\frac{\phi_{\text{SC}}}{N})\,.\label{eqn:Lspec}
\end{align}

For the Moore-Read (MR) FQH-SC at filling $\nu=1/m$, the Hilbert space is enlarged to $\Z_2^{(n)}\times\Z_{2m}$ due to the presence of Ising topological order in the parent MR FQH state \cite{moore-1991}. We label anyons in the MR FQH state by $a_r$, with $a\in\{1,\chi,\sigma\}$, where $1$ labels the vacuum, $\chi$ labels the Majorana fermion, and $\si$ labels the non-Abelian Ising anyon, respectively, and $r\in\Z_{4m}$ labels the charge sector. The tunneling spectrum can therefore contain contributions that depend on both neutral and charge fusion channels of the two defects, which we label as $(l_n,l)$. Here $l_n\in\{0,1\}\simeq\Z_{2}^{(n)}$ labels the neutral sector $1$ and $\chi$, respectively, and $l\in\Z_{2m}$ labels the charge sector. The superconducting phase couples only to the charge sector, while the neutral fusion channel contributes an additional channel-dependent phase shift. Focusing on the $\al_{\sigma_1}$ tunneling process \footnote{Here $\al_{\sigma_1}$ can be viewed as combining the charge sector zero mode $\al_{1_1}$ and the neutral sector zero mode $\al_{\si_0}$. We focus on tunneling of $\al_{\si_1}$ as the corresponding spectrum depends on both the neutral and charge fusion channels. Both $\al_{\si_1}$ and $\al_{1_1}$ transfer charge $e/(2m)$, and both are included where relevant below.}, the corresponding contribution to the tunneling spectrum is \cite{cao-2026}
\begin{align}
    E_{(l_n,l)}(\phi_{\text{SC}})&=2\abs{\la_{\si_1}}\cos(\frac{2\pi l}{2m}+\frac{2\pi l_n}{2}-\frac{\phi_{\text{SC}}}{2m})\,.\label{eqn:MRspec}
\end{align}
In both cases, the fusion channel can also be inferred from the phase dependence of the tunneling spectrum or the Josephson current, $I= 2e(\pa E/\pa\phi_{\text{SC}})$, where $e$ is the charge of an electron. 

\textit{QPC-induced fusion channel switching---}We now consider the effect of tunneling a quasiparticle $b$ across a single QPC. In the weak-tunneling, low-bias regime, tunneling across the QPC, with amplitude $\Ga$, can be treated perturbatively as single-quasiparticle tunneling events \cite{kane-1995,chamon-1997,bonderson-2006,fendley-2007,halperin-2011}. Since QPC tunneling is probabilistic, the resulting signature appears as random transitions between branches of the fractional Josephson spectrum, with transition rates controlled by $\abs{\Ga}^2$, analogous to the probabilistic effects seen in current FQH interferometers \cite{werkmeister-2025,samuelson-2026}. The transitions arise from the mutual linking between the worldline of $b$ and zero mode tunneling across the Josephson junction as shown in Fig.~\ref{fig:qpcj1}. Their mutual statistics change the phase of this zero mode tunneling operator and effectively shift the $X_2\times X_3$ fusion channel \cite{kapustin-2011,barkeshli-2013,barkeshli-2013-2,barkeshli-2019,shen-2019} (see details in Supplemental Material \cite{SM}):
\begin{align}
    \al_a^{(2)\da}\al_a^{(3)}\xrightarrow{b\text{ tunnels across QPC}}\pbr{\frac{\S_{ab}}{\S_{a0}}}^2\al_a^{(2)\da}\al_a^{(3)}\,.\label{eqn:zmalg}
\end{align}
Here $\S_{ab}$ is the modular $\S$-matrix encoding the linking between $a$ and $b$ \cite{verlinde-1988,moore-1989,kitaev-2006,bonderson-2007-thesis,fendley-2007-S,dong-2008} and $0$ denotes the vacuum sector. The squared statistical factor reflects braiding with the two counterpropagating edges along the FQH-SC interface. For fermionic states, $\S_{ab}$ is evaluated in the parity-resolved anyon description detailed in the Supplemental Material \cite{SM}.

\begin{figure}
    \centering
    \subfloat[]{%
        \includegraphics[width=0.48\columnwidth]{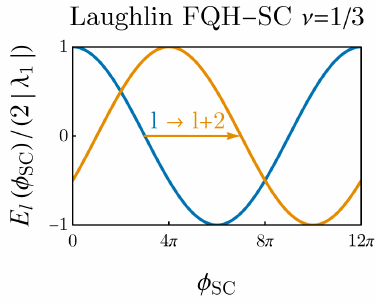}
        \label{fig:QPCL}
    }
    \subfloat[]{%
        \includegraphics[width=0.48\columnwidth]{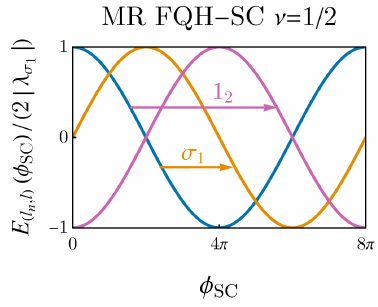}
        \label{fig:QPCMR}
    }
\caption{QPC-induced fusion-channel shifts. (a) In a Laughlin FQH-SC at $\nu=1/3$, tunneling a Laughlin quasihole shifts $l\to l+2$. (b) In an MR FQH-SC at $\nu=1/2$, tunneling $\si_1$ shifts the charge-sector fusion channel by one, while tunneling $1_2$ shifts it by two. The vertical axis shows the normalized Josephson energy. These fusion-channel shifts occur probabilistically as a result of quasiparticle tunneling at the QPC.}
\label{fig:QPC}
\end{figure}

We first illustrate this fusion-channel shift in a fermionic Laughlin FQH-SC at filling $\nu=1/m$. In the parity-resolved labeling used in Eq.~\eqref{eqn:zmalg}, the physical Laughlin quasihole of charge $e/m$ is represented by $b=2$. Its action on the $\Z_{2m}$ zero mode generator $\al_1$ is
\begin{align}
    \al^{(2)\da}_{1}\al^{(3)}_{1}\xrightarrow{b=2} e^{\frac{2\pi i}{2m}2}\al^{(2)\da}_{1}\al^{(3)}_{1}\,,\quad \ket{l}\xrightarrow{b=2}\ket{l+2}\,.\label{eqn:Lshift}
\end{align}
Comparing the statistical phase in Eq.~\eqref{eqn:Lshift} with the eigenvalues in Eq.~\eqref{eqn:zm_tunnel} shows that the QPC tunneling event shifts $l\to l+2$, thereby switching the Josephson spectrum from $E_l(\phi_{\text{SC}})$ to $E_{l+2}(\phi_{\text{SC}})$ \footnote{For more general Abelian FQH-SCs, the size of the shift ($l\to l+1$ or $l\to l+2$) depends on whether the defect Hilbert space is enlarged by fermion parity \cite{barkeshli-2013,barkeshli-2013-2}.}.

The fermionic MR FQH-SC at filling $\nu=1/m$ provides a non-Abelian example, in which a quasiparticle tunneling through the QPC can carry both electric charge and Ising topological charge \cite{moore-1991}. The minimally charged Ising anyon is $\sigma_1$, which carries charge $e/(2m)$. Because the neutral statistical factor contains $S_{\sigma\sigma}=0$, the neutral-sector contribution to zero mode tunneling vanishes, leaving the charge-sector contribution associated with $\al_{1_1}$. Equation \eqref{eqn:zmalg} then gives \footnote{Here the factor of $d_\si^2=2$ (the quantum dimension squared of the Ising anyon) rescales the overall tunneling amplitude but does not affect the fusion-channel dependence.}:
\begin{align}
    \al_{1_1}^{(2)\da}\al_{1_1}^{(3)}\xrightarrow{b=\si_1} 2e^{\frac{2\pi i}{2m}}\al_{1_1}^{(2)\da}\al_{1_1}^{(3)}\,,\quad \ket{l_n,l}\xrightarrow{b=\si_1} \ket{l_n,l+1}\,.
\end{align}
By contrast, tunneling of a Laughlin quasihole $1_2$ with charge $e/m$ shifts the same fusion channel by two,
\begin{align}
    \al_{1_1}^{(2)\da}\al_{1_1}^{(3)}\xrightarrow{b=1_2} e^{\frac{2\pi i}{2m}2}\al_{1_1}^{(2)\da}\al_{1_1}^{(3)}\,,\quad \ket{l_n,l}\xrightarrow{b=1_2} \ket{l_n,l+2}\,.
\end{align}
These different shifts are illustrated in Fig.~\ref{fig:QPC}. 

\textit{Interferometry for weak zero mode tunneling---}Having discussed how tunneling through a single QPC modifies the tunneling spectrum of the Josephson junction, we now consider the two-path Fabry-P\'{e}rot interference signal for a device of the type shown in Fig.~\ref{fig:FP}. As a reference point, we first consider the limit in which zero mode tunneling between $X_2$ and $X_3$ vanishes. In this limit, $X_1$ and $X_2$, located at the two ends of the same superconducting region, are in a definite fusion channel. In the interferometer, a quasiparticle can tunnel through either QPC, and the two tunneling paths interfere. The interference phase contains two contributions: the Aharonov-Bohm (AB) phase from the enclosed magnetic flux and the statistical phase from braiding around the anyons inside the interferometer. 

The two-path interference signal appears directly in the low-bias tunneling conductance. For a generic FQH state, the longitudinal conductance takes the form \cite{chamon-1997,fradkin-1998,bonderson-2006,stern-2006,stern-2010,bonderson-2009,bishara-2009,halperin-2011},
\begin{equation}
    \begin{split}
    \si_{xx}\propto  \,&\abs{\Ga_1}^2+\abs{\Ga_2}^2\\
    &+2\Re(\Ga_1^\ast\Ga_2 e^{i\vp_{\text{AB}}}\ev{W_{N_{\text{qh}},b}}{\Psi})\,,
    \end{split}\label{eqn:general_fp}
\end{equation}
where $\Ga_i$ is the quasiparticle tunneling amplitude at QPC $i$, $\vp_{\text{AB}}$ is the AB phase, $N_{qh}$ is the number of fundamental quasiholes in the interferometer, and $\ket{\Psi}$ denotes the state of the topological charge enclosed by the interferometer. 
Unlike tunneling events through one QPC, $W_{N_{\text{qh}},b}$ is the closed Wilson loop associated with the two-path interference process and encircles the total topological charge within the interferometer. If the total enclosed topological charge is $c$, this expectation value reduces to monodromy between $c$ and $b$ \cite{bonderson-2006,stern-2010}, which encodes full braiding between them:
\begin{align}
    \ev{W_{N_{\text{qh}},b}}{\Psi_c}=M_{cb}\,,\quad M_{cb}\equiv\frac{\S_{cb}\S_{00}}{\S_{0c}\S_{0b}}\,.
\end{align}

In the weak zero mode tunneling limit, the pair of defects $X_1$ and $X_2$ at the ends of SC1 has a definite fusion channel and therefore acts as a trapped topological charge inside the interferometer. Equation \eqref{eqn:general_fp} then reduces the measurement to standard FQH Fabry-P\'{e}rot interferometry, with the defect fusion channel entering the interference signal through its monodromy with the tunneling quasiparticle. For a fermionic Laughlin FQH-SC, different defect fusion channels produce distinct statistical phase shifts. For a fermionic MR FQH-SC, both the neutral and charge components contribute phase shifts, while the usual Moore-Read even-odd effect is preserved: interference vanishes when an odd number of Ising anyons $\si$ is enclosed and is restored when that number is even \cite{bonderson-2006,stern-2006,bishara-2009}.

\textit{Interferometry for strong zero mode tunneling---}We now consider the strong zero mode tunneling regime of this device, in which zero mode tunneling across the Josephson junction selects a definite $X_2 \times X_3$ fusion channel. A state with a definite $X_2\times X_3$ fusion channel is generally a superposition of $X_1 \times X_2$ fusion channels. 
Since the interferometer loop operator is diagonal in the $X_1\times X_2$ basis, the interference term becomes a weighted sum over monodromy phases arising from braiding in these different fusion channels. For the states considered here, this superposition produces destructive interference, distinguishing the strong from the weak zero mode tunneling regime. Explicitly, the change of basis between  $X_2\times X_3$ and $X_1\times X_2$ fusion bases is implemented by a defect $F$-move, whose matrix elements are defect $F$-symbols \cite{kitaev-2006,nayak-2008,barkeshli-2019}. Indeed, for the fermionic Laughlin FQH-SC at filling $\nu=1/m$, the defect $F$-move gives an equal-weight sum over the $\Z_{2m}$ $X_1 \times X_2$ fusion channels, and the corresponding monodromy phases cancel by orthogonality \cite{SM}. Consequently,
\begin{align}
    \si_{xx}\propto \abs{\Ga_1}^2+\abs{\Ga_2}^2\,+\ldots\,,\label{eqn:noint}
\end{align}
where the ellipsis denotes perturbative corrections in the strong zero mode tunneling limit.

This result provides a particularly sharp signature of the non-Abelian fusion structure of the superconducting defects. Such a signature is absent in an ordinary Abelian FQH state, as the fusion space of Abelian anyons is one-dimensional and therefore does not support a nontrivial change of fusion basis. By contrast, superconducting defects in an Abelian FQH state possess a multidimensional fusion space, allowing the interference contributions from different fusion channels to cancel exactly. Consequently, the interference term vanishes identically in the topological limit, rather than merely acquiring a channel-dependent phase shift or a reduced amplitude.

Although Eq.~\eqref{eqn:noint} has the same conductance form as the well-known Moore–Read even-odd effect, the mechanisms are distinct. In the MR case, the interference term vanishes for an odd number of enclosed $\si$ anyons because $M_{\si\si}=0$ \cite{stern-2006,bonderson-2006}. However, in our case each Abelian monodromy is nonzero, and the exact cancellation instead results from summing over the $X_1\times X_2$ fusion channels related by the defect $F$-move. This cancellation occurs for Abelian FQH-SCs in general, irrespective of whether fermion parity enlarges the zero mode Hilbert space.

The exact cancellation also extends to the fermionic MR FQH-SC. In this case, the defect $F$-move factorizes into neutral and charge sectors, transforming a definite $X_2\times X_3$ fusion channel labeled by $(l_n,l)$ into an equal-weight superposition of $X_1\times X_2$ fusion channels labeled by $(k_n,k)$. Accordingly, the interferometric cross term factorizes into neutral- and charge-sector sums. For a topologically nontrivial tunneling quasiparticle, at least one of these sums vanishes by orthogonality, and hence the full cross term cancels \cite{SM}. Thus, the strong-tunneling disappearance of interference occurs for defects arising from both Abelian and non-Abelian parent FQH states. 

\textit{Discussion---}We have proposed two related configurations based on an FQH-SC Josephson junction coupled to QPCs.  In the single-QPC configuration, quasiparticle tunneling switches defect fusion channels; in the two-QPC Fabry–P\'{e}rot interferometer, interference probes them. In the strong zero mode tunneling regime, the defect $F$-move produces exact cancellation of the interference term, providing a sharp signature of the defects’ non-Abelian fusion structure even for a parent Abelian FQH state. 

Our proposal is has many similarities to that of  Ref.~\cite{barkeshli-2014}, which considered bilayer FQH states with a pair of interlayer tunneling defects; these play the same role as the defects at the ends of the SC strips in our setup, yielding a pair of PZMs, which the authors proposed to probe using inteferometric measurements. In contrast to our setup, Ref.~\cite{barkeshli-2014} proposes using two pairs of QPCs. In the weak tunneling limit, closed loops of anyons through these QPCs amount to two Wilson loops; the linking of these loops manifests in a current-noise cross-correlation. In contrast, we employ a single pair of QPCs, which yield one Wilson loop. The other Wilson loop corresponds to the anyon tunneling operator itself. We proposed to measure the linking of these loops through destructive interference in the linear conductance. Thus, while both our proposal and that of Ref.~\cite{barkeshli-2014} probe PZMs using interferometry, they differ in the precise setup. Moreover, the switching of the fusion channel monitored through the fractional Josephson spectrum proposed here is not discussed in Ref.~\cite{barkeshli-2014}.

In an experimental device, finite defect separations generate residual zero mode couplings absent in the idealized limits. As shown in the End Matter, weak but finite tunneling across the junction changes the magnitude and phase of the interference at second order. These leading corrections are independent of $\phi_{\rm SC}$, while higher-order terms acquire Josephson-phase dependence. In the strong zero mode tunneling limit, residual zero mode hybridization within a superconducting segment restores interference only perturbatively: the leading contribution is second order for the fermionic Laughlin FQH-SC and first order for the MR FQH-SC when an even number of Ising anyons is enclosed. Thus, when the residual hybridization is weak, the restored interference remains strongly suppressed relative to the weak-limit signal. Its dependence on $\phi_{\rm SC}$ generally contains all harmonics allowed by the fractional Josephson periodicity away from branch crossings.

The topological mechanism is not restricted to Landau-level-based FQH states. It applies more generally to charge-enriched topological orders whose edges can be proximitized by superconductors, including FQAH states \cite{haldane-1988,neupert-2011,tang-2011,sheng-2011,regnault-2011} and fractional quantum spin Hall states \cite{levin-2009}. FQAH platforms are especially promising for interferometry because they realize FQH-like topological order without requiring an external magnetic field. Their operation therefore need not rely on sweeping a magnetic field, potentially reducing Coulomb-driven changes of the interferometer area that can obscure anyonic statistical signatures \cite{halperin-2011}. Gate voltages and QPCs can instead adjust the edge trajectories and quasiparticle-tunneling amplitudes. 

Additional internal symmetries may further enrich the defect fusion structure, as in two-component FQH states with $SU(2)$ spin or layer-pseudospin symmetry \cite{halperin-1984,balatsky-1991,wen-1992,lee-1994,read-2000,lopez-2001}, their non-Abelian spin-singlet generalizations \cite{ardonne-2001}, and graphene-based systems with spin and valley degeneracies \cite{nomura-2006}. Determining how these structures modify the Josephson and interferometric signatures provides an interesting direction for future work.

\textit{Acknowledgments---}We thank John McGreevy for motivating the setup to us, Da-Chuan Lu and Taige Wang for discussions of their unpublished work, and Maissam Barkeshli for discussions regarding Ref.~\cite{barkeshli-2014}. This work was supported in part by the U.S. National Science Foundation through the grant DMR 2225920 at the University of Illinois (JC and EF) and by a MURI ONR grant Award No. N00014-22-1-2764 (AK). 
\bibliography{MR}
\clearpage
\appendix
\onecolumngrid
\begin{center}
    {\large\bfseries End Matter}
\end{center}
\twocolumngrid
\setcounter{equation}{0}
\renewcommand{\theequation}{A\arabic{equation}}
\textit{Interferometric signals for the Laughlin and Moore-Read states---}In the main text, we discussed the interferometric signatures in the ideal weak and strong zero mode tunneling limits. In an experimental device, the separation between all defects is finite, so neither zero mode hybridization within a superconducting segment nor zero mode tunneling across the junction vanishes exactly \cite{chen-2016,teixeira-2022}. We therefore focus on the zero modes hosted at three defects $X_1,~X_2,~X_3$ \footnote{Since conservation of fermion parity implies that the fusion channel of the four defects is always fixed \cite{lindner-2012}, the fusion channels involving $X_4$ are not independent.}. We then consider the Hamiltonian $H=H_{12}+H_{23}$, where
\begin{align}
    H_{12}={\textstyle\sum_{a}} \ka_a \al_{a}^{(1)\da}\al_{a}^{(2)}+\hc
\end{align}
Here $\ka_a$ is the hybridization amplitude between zero modes $\al_a^{(1)}$ and $\al_a^{(2)}$ within SC1 \cite{chen-2016,teixeira-2022}, and $H_{23}$ is the zero mode tunneling Hamiltonian in Eq.~\eqref{eqn:hamiltonian}, with amplitude $\la_a$. The limits considered below are determined by the relative magnitudes of $\abs{\ka_a}$ and $\abs{\la_a}$. In particular, the limits of (i) no zero mode tunneling, (ii)  weak zero mode tunneling, and (iii) strong zero mode tunneling correspond to, respectively, (i) $\abs{\la_a}=0$ with finite $\abs{\ka_a}$, (ii) $\abs{\la_a}\ll\abs{\ka_a}$, and (iii) $\abs{\la_a}\gg\abs{\ka_a}$.

\textit{(i) No zero mode tunneling limit---}We first consider the limit $\abs{\la_a}=0$ with finite $\abs{\ka_a}$, which provides the unperturbed interferometric response. In this limit, an eigenstate of $H_{12}$ corresponds to a definite fusion channel of $X_1\times X_2$. For a fermionic Laughlin FQH-SC at filling $\nu=1/m$, we label the $X_1\times X_2$ fusion channel by $k\in\Z_{2m}$. The longitudinal conductance is $\si_{xx}(k)=\si_{xx}^{(0)}+\si_{xx}^{\rm int}(k)$, where $\si_{xx}^{(0)}\propto \abs{\Ga_1}^2+\abs{\Ga_2}^2$ arises from tunneling independently at the two QPCs and $\si_{xx}^{\rm int}(k)$ denotes the interference term:
\begin{align}
    \si_{xx}^{\rm int}(k)&\propto 2\abs{\Ga_1}\abs{\Ga_2} \cos\Th_k\,,\\
    \Th_k&\equiv\frac{2\pi}{m}\pbr{\frac{\Phi}{\Phi_0}+N_{\text{qh}}+k}\,,\label{eqn:LFP}
\end{align}
where $\Th_k$ denotes the interference phase and $\Phi/\Phi_0$ is the flux enclosed by the interfering paths measured in units of flux quantum $\Phi_0=2\pi/e$ and $N_{\rm qh}$ is the number of enclosed Laughlin quasiholes \footnote{Although defect fusion channels labeled by $k$ and $k+m$ are distinct with opposite fermion parity, they give the same interference signal because the physical quasihole has identical monodromy with channels that differ by fusion with an electron.}.

For a fermionic Moore-Read FQH-SC, we label the $X_1\times X_2$ fusion channel by $(k_n,k)$, where $k_n\in\Z_2^{(n)}$ and $k\in\Z_{2m}$. When the number $N_\sigma$ of enclosed Ising anyons is odd, the interference term vanishes $\si_{xx}\propto\si_{xx}^{(0)}$.
When $N_\sigma$ is even, the interference is restored and
\begin{align}
    &\si_{xx}^{\rm int}(k_n,k)\propto2\abs{\Ga_1}\abs{\Ga_2}\cos\Th_{(k_n,k)}\\
    &\Th_{(k_n,k)}\equiv\sbr{\frac{2\pi}{2}(N_\chi+k_n)+\frac{2\pi}{4m}\pbr{2\frac{\Phi}{\Phi_0}+N_\si+2k}}\,,
\end{align}
where $N_\chi\in\Z_2$ is the fusion channel of the enclosed Ising anyons.

\textit{(ii) Weak zero mode tunneling limit---}We then consider the limit where the zero mode tunneling between $X_2$ and $X_3$ is weak but finite, corresponding to $\abs{\la_a}\ll\abs{\ka_a}$. For a fermionic Laughlin FQH-SC at filling $\nu=1/m$, the unperturbed basis is $\ket{k}$ with the $H_{12}$ spectrum:
\begin{align}
    H_{12}\ket{k}=E_k^{(12)}\ket{k}\equiv2\abs{\ka_1}\cos(\frac{\pi k}{m}+\theta_{12})\ket{k}\,,
\end{align}
where $\theta_{12}$ is the microscopic zero mode hybridization phase, which depends on the chemical potential of the superconducting segment and the separation between defects $X_1$ and $X_2$ \cite{chen-2016}. Away from crossings of the $H_{12}$ spectrum, the eigenstates $\ket{k}$ are nondegenerate and $H_{23}$ can be treated using ordinary perturbation theory. Evaluating the matrix elements $\mel{k'}{H_{23}}{k}$ using the defect $F$-symbols gives nonzero contributions only for $k'=k\pm1$. The perturbed state is then
\begin{align}
    \ket{\tilde{k}}\propto \ket{k}+\frac{\la_1}{\De_{k,+}}\ket{k+1}+\frac{\la_1^\ast}{\De_{k,-}}\ket{k-1}+\O(\la^2_1)\,,\label{eqn:Lpert}
\end{align}
where $\De_{k,\pm}\equiv E_k^{(12)}-E_{k\pm 1}^{(12)}$ is the energy difference between $\ket{k}$ and $\ket{k\pm 1}$. After normalizing the perturbed state in Eq. ~\eqref{eqn:Lpert} and evaluating the interferometer Wilson loop operator, we obtain, to quadratic order in $\la_1$,
\begin{equation}
    \begin{split}
        \si_{xx}^{\rm int}(\tilde{k})&\propto2\abs{\Ga_1}\abs{\Ga_2}\bigg[1-2\abs{\la_1}^2\sin^2\pbr{\frac{\pi}{m}}\\
        &\pbr{\frac{1}{\De_{k,+}^2}+\frac{1}{\De_{k,-}^2}}\bigg]\cos(\Th_k+\de\Th_k)\,,
    \end{split}
\end{equation}
The factor in square brackets reduces the magnitude of the interference, while zero mode tunneling shifts its phase by $\de\Th_k$, where
\begin{align}
    \de\Th_k=\abs{\la_1}^2\sin(\frac{2\pi}{m})\pbr{\frac{1}{\De_{k,+}^2}-\frac{1}{\De_{k,-}^2}}.
\end{align}

For a fermionic MR FQH-SC, two zero mode tunneling processes contribute to the perturbative correction.  In the $X_1\times X_2$ fusion basis, the tunneling operator $\al_{\si_1}^{(2)\da}\al_{\si_1}^{(3)}$ connects $\ket{k_n,k}$ to $\ket{k_n+1,k\pm 1}$ through the defect $F$-move. By contrast, the charge-sector $\Z_{2m}$ PZM tunneling operator $\al_{1_1}^{(2)\da}\al_{1_1}^{(3)}$ connects $\ket{k_n,k}$ to $\ket{k_n,k\pm1}$ and changes only the charge sector.  We define the corresponding excitation energies as $\De_{\si_1,\pm}^{(12)}=E^{(12)}_{(k_n,k)}-E^{(12)}_{(k_n+1,k\pm1)}$ and $\De_{1_1,\pm}^{(12)}=E^{(12)}_{(k_n,k)}-E^{(12)}_{(k_n,k\pm1)}$. For simplicity, we take the unperturbed zero mode hybridization \(H_{12}\) to be dominated by the \(\alpha_{\sigma_1}\) process, whose spectrum has the same fusion-channel dependence as Eq.~\eqref{eqn:MRspec}:
\begin{align}
    E^{(12)}_{(k_n,k)}=2\abs{\ka_{\si_1}}\cos(k_n\pi+\frac{\pi k}{m}+\th_{12})\,.
\end{align}
For even $N_\si$, the interference term to second order in zero mode tunneling amplitudes $\la_{\si_1}$ and $\la_{1_1}$ is
\begin{widetext}
\begin{equation}
    \begin{split}
        \si_{xx}^{\rm int}(\tilde{k}_n,\tilde{k})\propto2\abs{\Ga_1}\abs{\Ga_2}\cos\pbr{\Th_{(k_n,k)}+\de\Th_{(k_n,k)}}&\bigg\{1-2\abs{\la_{\si_1}}^2\cos^2\pbr{\frac{\pi}{2m}}\sbr{\pbr{\De_{\si_1,+}^{(12)}}^{-2}+\pbr{\De_{\si_1,-}^{(12)}}^{-2}}\\
        &-2\abs{\la_{1_1}}^2\sin^2\pbr{\frac{\pi}{2m}}\sbr{\pbr{\De_{1_1,+}^{(12)}}^{-2}+\pbr{\De_{1_1,-}^{(12)}}^{-2}}\bigg\}\,,
    \end{split}
\end{equation}
\end{widetext}
where the phase shift $\de\Th_{(k_n,k)}$ is
\begin{equation}
    \begin{split}
        \de\Th_{(k_n,k)}&=\abs{\la_{\si_1}}^2\sin(\frac{\pi}{m})\sbr{\pbr{\De_{\si_1,-}^{(12)}}^{-2}-\pbr{\De_{\si_1,+}^{(12)}}^{-2}}\\
        &+\abs{\la_{1_1}}^2\sin(\frac{\pi}{m})\sbr{\pbr{\De_{1_1,+}^{(12)}}^{-2}-\pbr{\De_{1_1,-}^{(12)}}^{-2}}\,.
    \end{split}
\end{equation}
For odd $N_\si$, the interference term remains zero.

In summary, for both the Laughlin and MR FQH-SCs, these corrections modify the magnitude and phase of the interference term. At this order, the corrections are independent of $\phi_{\rm SC}$; dependence on $\phi_{\rm SC}$ enters at higher orders.

\textit{(iii) Strong zero mode tunneling limit---}In the strong zero mode tunneling limit, to leading order, the interference term vanishes as in Eq.~\eqref{eqn:noint}. When the hybridization between defects $X_1$ and $X_2$ is finite, we take the zero mode tunneling Hamiltonian in Eq.~\eqref{eqn:hamiltonian} as the unperturbed Hamiltonian, and $H_{12}$ as the perturbation. For a fermionic Laughlin FQH-SC at filling $\nu=1/m$, the zero mode hybridization operator $\al_1^{(1)\da}\al_1^{(2)}$ acts as a shift operator $l\to l\pm 1$ through the defect $F$-move. The perturbed state is then
\begin{equation}
    \begin{split}
        &\ket{\tilde{l}}\propto\ket{l}+\frac{\ka_{1}}{\De_{l,+1}}\ket{l+1}+\frac{\ka_{1}^\ast}{\De_{l,-1}}\ket{l-1}\\
        &+\frac{\ka_1^2}{\De_{l,+1}\De_{l,+2}}\ket{l+2}+\frac{\pbr{\ka_1^\ast}^2}{\De_{l,-1}\De_{l,-2}}\ket{l-2}+\O(\ka_1^3)\,,
    \end{split}
\end{equation}
where $\De_{l,\pm n}\equiv E_l(\phi_{\rm SC})-E_{l\pm n}(\phi_{\rm SC})$ is the energy difference between $\ket{l}$ and $\ket{l\pm n}$. After normalizing the perturbed state and evaluating the interferometer loop operator, we obtain, to quadratic order in $\ka_1$,
\begin{equation}
    \begin{split}
        \si_{xx}^{\rm int}(\tilde{l})&\propto 2\abs{\Ga_1}\abs{\Ga_2}\abs{\ka_1}^2\cos(\Th_0-2\th_{12})\\
        &\pbr{\frac{1}{\De_{l,+1}\De_{l,+2}}+\frac{1}{\De_{l,-1}\De_{l,-2}}+\frac{1}{\De_{l,+1}\De_{l,-1}}}\,,
    \end{split}
\end{equation}
where $\Th_0\equiv\Th_{k=0}$ is the unperturbed interference phase. This means that finite zero mode hybridization produces a leading correction to the strong-tunneling interference at second order in $\ka_1$, because two hybridization processes are required to match the fusion-channel shift generated by the quasihole loop. This correction restores a small interference contribution whose magnitude and phase depend on the hybridization amplitude, the energy gaps between fusion channel branches of the Josephson spectrum, and the microscopic hybridization phase $\th_{12}$.

For a fermionic MR FQH-SC, we focus on the regime in which the $\al_{\si_1}$ tunneling process dominates $H_{23}$. We label the corresponding eigenstates by $\ket{l_n,l}$, with spectrum given by Eq.~\eqref{eqn:MRspec}. Hybridization through $\al_{\si_1}$, with amplitude $\ka_{\si_1}$, changes the state from $\ket{l_n,l}$ to $\ket{l_n+1,l\pm 1}$. The Wilson loop operator $W_{N_{qh},\si_1}$ is diagonal in the $X_1\times X_2$ basis, but it is represented, after the defect $F$-move, as
\begin{align}
    W_{N_{qh},\si_1}\ket{l_n,l}\propto\ket{l_n+1,l+1}\,,
\end{align}
in the $X_2\times X_3$ basis. Since the $\al_{\si_1}$ hybridization process changes the neutral sector of the fusion channel by one, $l_n\to l_n+1$, the leading correction to the interference is first order in $\ka_{\si_1}$. For even $N_\si$, the interference term is first order in perturbation theory:
\begin{equation}
    \begin{split}
        \si_{xx}^{\rm int}(\tilde{l}_n,\tilde{l})\propto 2\abs{\Ga_1}\abs{\Ga_2}&\abs{\ka_{\si_1}}\cos(\Th_{(0,0)}-\th_{12})\\
        &\pbr{\frac{1}{\De_{\si_1,+}^{(23)}}+\frac{1}{\De_{\si_1,-}^{(23)}}}\,,
    \end{split}
\end{equation}
where $\Th_{(0,0)}\equiv\Th_{(k_n=0,k=0)}$ is the unperturbed interference phase and $\De_{\si_1,\pm}^{(23)}=E_{(l_n,l)}-E_{(l_n+1,l\pm1)}$ is the energy difference between $\ket{l_n,l}$ and $\ket{l_n+1,l\pm 1}$. For odd $N_\si$, the interference remains zero, at least to this  order in the strong zero mode tunneling expansion.

The energy differences $\De_{l,\pm n}$ and $\De_{\si_1,\pm}^{(23)}$ depend explicitly on the superconducting phase difference $\phi_{\rm SC}$ through the fractional Josephson spectrum. The leading interference corrections in both the Laughlin and MR cases are therefore tunable by $\phi_{\rm SC}$. Because these corrections contain inverse powers and products of the energy differences, their dependence on $\phi_{\rm SC}$ is generally nonsinusoidal and contains all harmonics allowed by the fractional Josephson periodicity. These perturbative expressions apply away from branch crossings, where the relevant energy differences remain nonzero.

\clearpage
\onecolumngrid
\begin{center}
    {\large\bfseries Supplemental Material for Interferometric Signatures of Zero Modes in Fractional Quantum Hall-Superconductor Heterostructures}
\end{center}
\setcounter{equation}{0}
\counterwithout{equation}{section}
\renewcommand{\theequation}{S\arabic{equation}}
\setcounter{section}{0}
\renewcommand{\thesection}{\Roman{section}}
\setcounter{subsection}{0}
\renewcommand{\thesubsection}{\thesection.\Alph{subsection}}
\setcounter{secnumdepth}{2}
\section{Fermion parity and minimal modular extension}
The zero mode algebra in the main text applies most directly to bosonic topological orders. For fermionic topological orders, including electronic FQH states, the defect Hilbert space may contain an additional contribution associated with fermion parity \cite{kitaev-2001,fidkowski-2010,fidkowski-2011,turner-2011,barkeshli-2013}. We incorporate this contribution by embedding the fermionic theory into a minimal modular extension \cite{bruillard-2017,aasen-2022,kobayashi-2022}. The extension contains the anyons and physical electron of the original theory together with fermion-parity fluxes and provides the nondegenerate modular $\S$-matrix used in the zero mode algebra. The minimal modular extension serves only as a technical tool; the quasiparticles that tunnel through the QPC remain deconfined anyons of the parent FQH state.
\subsection{Fermionic Laughlin FQH-SC}
The fermionic Laughlin state at filling $\nu=1/m$, with $m$ odd, has $m$ topologically distinct quasiparticle sectors. To incorporate fermion parity, we use the minimal modular extension $U(1)_{4m}$. The sectors of the extended theory are labeled by $r\in\Z_{4m}$, carry electric charge $q_r=re/(2m)$ and have modular $\S$-matrix
\begin{align}
    \S_{rs}^{U(1)_{4m}}=\frac{1}{\sqrt{4m}}e^{\frac{2\pi i}{4m} rs}\,.
\end{align}
The physical electron is the sector $r_\psi=2m$. The deconfined quasiparticles of the parent Laughlin state correspond to the even sectors $r=2j$. After identifying sectors that differ by fusion with the physical electron, these quasiparticles are labeled by $j\in\Z_m$. The corresponding FQH-SC has a $\Z_{2m}$ zero mode structure. 

In the main text, we label the $X_1\times X_2$ fusion channel of two superconducting defects by $k\in\Z_{2m}$, with fermion parity $(-1)^F\ket{k}=(-1)^k \ket{k}$. Fusion with the physical electron shifts $k\ra k+m$. Because $m$ is odd, the channels $k$ and $k+m$ have opposite fermion parity and are therefore distinct channels of the defect Hilbert space. Nevertheless, the physical electron is transparent to the deconfined quasiparticles of the parent Laughlin state. These quasiparticles consequently have identical monodromy with $k$ and $k+m$. This explains why the two channels give the same Fabry–P\'{e}rot signal in the main text, even though they have opposite fermion parity.

\subsection{Fermionic Moore-Read FQH-SC}
The fermionic Moore–Read state at filling fraction $\nu=1/m$, with $m$ even, consists of an Ising neutral sector and a $U(1)_{4m}$ charge sector. We use the minimal modular extension $\text{Ising}\,\times U(1)_{4m}$. The sectors are labeled by $a_r$, where $a\in\cbr{1,\si,\chi}$ denotes the Ising sector and $r\in\Z_{4m}$ denotes the charge sector. The modular $\S$-matrix factorizes \cite{fendley-2007-S,dong-2008} as
\begin{align}
    \S_{a_rb_s}^{\text{MR}}=\S_{ab}^{\text{Ising}}\S_{rs}^{U(1)_{4m}}\,,
\end{align}
where the modular $\S$-matrix for the Ising neutral sector in the $\cbr{1,\si,\chi}$ basis is
\begin{align}
    \S_{ab}^{\text{Ising}}\equiv \frac{1}{2}\mqty(1&\sqrt{2}&1\\\sqrt{2}&0&-\sqrt{2}\\1&-\sqrt{2}&1)\,.
\end{align}
The physical electron is $\psi=\chi_{2m}$. In the physical anyon theory, locality with respect to the electron requires the Abelian anyons $1_r$ and $\chi_r$ to have even charge-sector labels $r$, whereas the Ising anyon $\si_r$ has odd charge-sector label $r$.

The fusion channels relevant to the FQH-SC are labeled by $(l_n,l)\in\Z_2^{(n)}\times\Z_{2m}$, where the values $l_n=0,1$ correspond to the neutral fusion channel $1$ and $\chi$, respectively, and $l$ labels the charge sector.
\section{QPC tunneling and zero mode algebra}
Having established the modular data needed for fermionic FQH-SCs, we now provide a brief derivation of the zero mode algebra used in the main text (see Ref.~\cite{cao-2026} for details). We first identify the QPC and zero mode processes in the physical device geometry and then fold the counterpropagating edges to relate their crossing to the half-linking of topological lines in a system with gapped boundaries.
\begin{figure}
    \centering
    \includegraphics[width=0.40\columnwidth]{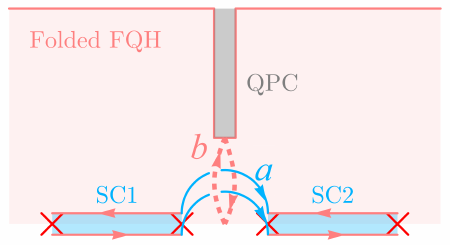}
    \caption{Folded representations of the QPC and zero mode processes. After folding the counterpropagating edges, the superconducting regions become gapped boundaries and the two processes form a half-linking configuration. The two components of the folded theory each contribute $\S_{ab}/\S_{a0}$, yielding the squared factor in Eq.~\eqref{eqn:zmalg}.}
    \label{fig:qpcj2}
\end{figure}

Figure \ref{fig:qpcj2} shows the two processes in the physical FQH-SC geometry. The $b$-labeled QPC process transfers an anyon $b$ across the constriction, while the $a$-labeled zero mode process connects the adjacent endpoints of the two superconducting regions. Projecting these processes onto the low-energy defect Hilbert space gives the QPC operator $U_b$ and the zero mode operator $\al_a$, respectively \cite{barkeshli-2013,cao-2026}.

To expose the algebra between these operators, we fold the FQH-SC heterostructure across the FQH-SC interfaces, as in Fig.~\ref{fig:qpcj2}. After folding one side of each FQH-SC interface, the two counterpropagating edge modes are represented as the two sectors of the folded topological theory, while the superconducting regions become gapped boundaries. The zero mode and QPC tunneling processes are then represented by crossing open topological lines labeled by $a$ and $b$, respectively. This crossing takes the form of the half-linking process for a system with gapped boundaries \cite{kapustin-2011,shen-2019,cao-2026}. 

For the FQH-SC boundary considered here, the half-linking factor separates into contributions from the two components of the folded theory. Each component contributes $\S_{ab}/\S_{a0}$, where $\S_{ab}$ is an element of the modular $\S$-matrix of the parent FQH state and $0$ denotes the vacuum. Denoting the action of the QPC-mediated anyon tunneling on $\al_a$ by $\U_b$, the half-linking relation gives 
\begin{align}
    \U_b[\al_a^{(2)\da}\al_a^{(3)}]=\pbr{\frac{\S_{ab}}{\S_{a0}}}^2\al_a^{(2)\da}\al_a^{(3)}\,.\label{eqn:smzmalg}
\end{align}
When $U_b$ is invertible, this action reduces to ordinary conjugation, 
\begin{align}
    \U_b[\al_a^{(2)\da}\al_a^{(3)}]=U_b^\da \al_a^{(2)\da}\al_a^{(3)}U_b\,.
\end{align}
Equation \eqref{eqn:smzmalg} reproduces the zero mode algebra used in the main text. For fermionic FQH-SCs, the modular data in Eq.~\eqref{eqn:smzmalg} are evaluated in the minimal modular extension described previously, while $b$ remains a deconfined quasiparticle of the parent FQH state.

\subsection{Fermionic Laughlin FQH-SC}
For the fermionic Laughlin state at filling fraction $\nu=1/m$, with $m$ odd, the zero mode generator is $\al_{1}$. The minimally charged Laughlin quasihole is represented by the sector $b=2$ in the $U(1)_{4m}$ minimal modular extension. Using the modular $\S$-matrix, we find
\begin{align}
    \pbr{\frac{\S_{12}}{\S_{10}}}^{2}=e^{\frac{2\pi i}{2m}2}\,.
\end{align}
Because $U_{2}$ is invertible, the zero mode algebra becomes
\begin{align}
    U_{2}^{\dagger}\al_1^{(2)\da}\al_1^{(3)}U_{2}=\pbr{\frac{\S_{12}}{\S_{10}}}^{2}\al_1^{(2)\da}\al_1^{(3)}=e^{\frac{2\pi i}{2m}2}\al_1^{(2)\da}\al_1^{(3)}\,.
\end{align}

The eigenvalue of $\al_{1}$ distinguishes the $\Z_{2m}$ fusion channels. The QPC tunneling event therefore
shifts the $X_{2}\times X_{3}$ fusion channel $l\ra l+2$. Since the shift is even, the fermion parity is preserved.

For the Jain and Halperin FQH-SCs considered in the main text, fermion parity does not enlarge the defect Hilbert space \cite{barkeshli-2013}. Tunneling the minimally charged quasiparticle therefore shifts the corresponding fusion-channel label by one: $l\ra l+1$.

\subsection{Fermionic Moore-Read FQH-SC}
For the fermionic MR state, the modular $\S$-matrix factorizes into its neutral Ising and charged $U(1)_{4m}$ components. The fusion channels are labeled by $(l_{n},l)\in\Z_{2}^{(n)}\times\Z_{2m}$, where $l_{n}$ labels the neutral fusion channel and $l$ labels the charge fusion channel.

Tunneling the minimally charged quasiparticle $\sigma_{1}$ suppresses the neutral-sector zero mode contribution because $\S_{\si\si}^{\text{Ising}}=0$. Consequently, $\U_{\si_1}[\al_{\si_1}]=0$.
The leading surviving contribution therefore comes from the charge-sector zero mode $\al_{1_{1}}$. Applying the zero mode algebra gives
\begin{align}
    \U_{\si_1}[\al_{1_1}^{(2)\da}\al_{1_1}^{(3)}]=2e^{\frac{2\pi i}{2m}}\al_{1_1}^{(2)\da}\al_{1_1}^{(3)}\,.
\end{align}
The factor $d_{\sigma}^{2}=2$, corresponding to the squared quantum dimension of the Ising anyon, is independent of the fusion-channel label and therefore does not affect the channel shift. The phase gives $(l_{n},l)\ra(l_{n},l+1)$. For tunneling of the Abelian quasihole $1_{2}$, the action reduces to ordinary conjugation,
\begin{align}
    U_{1_{2}}^{\da}\al_{1_1}^{(2)\da}\al_{1_1}^{(3)}U_{1_{2}}=e^{\frac{2\pi i}{2m}2}\al_{1_1}^{(2)\da}\al_{1_1}^{(3)}\,,
\end{align}
and the corresponding fusion-channel shift is $(l_{n},l)\ra(l_{n},l+2)$. These results give the QPC-induced fusion-channel shifts used in the main text.

\section{Fusion-basis transformation and interferometric response} \label{app:FP}
In this section, we derive the interferometric response when the Josephson junction and the Fabry-P\'{e}rot interferometer probe different fusion bases of the four-defect Hilbert space. Because the conductance depends on the expectation value of the interferometer loop operator, expressing the defect state in the interferometer basis leads to a weighted sum over fusion channels.

We denote the four defects by $X_1$, $X_2$, $X_3$, and $X_4$. In the strong zero mode tunneling regime, the zero mode tunneling between SC1 and SC2 pins the fusion channel between defects $X_2$ and $X_3$. A state in this basis is written as
\begin{align}
    \ket{X_2\times X_3=a, X_1\times X_4=\bar{a}}\,.
\end{align}
Here $\bar{a}$ is the conjugate of $a$ with fusion rule $a\times\bar{a}\ni0$ since we assume that the total fusion channel of the four defects is fixed to be the vacuum sector $0$. By contrast, the Fabry-P\'{e}rot interferometer measures the fusion channel between defects $X_1$ and $X_2$, because the loop operator associated with the tunneling quasiparticle $W_{N_{\text{qh},b}}$ encircles the region containing these defects, and it is diagonal in the basis
\begin{align}
    \ket{X_1\times X_2=c,~X_3\times X_4=\bar{c}}\,.
\end{align}
The two bases are related by an $F$-move
\begin{align}
    \ket{X_2\times X_3=a,~X_1\times X_4=\bar{a}}=\sum_{c} \sbr{F_{XXX}^X}_{ac}\ket{X_1\times X_2=c,~X_3\times X_4=\bar{c}}\,,
\end{align}
where $X$ denotes the common defect type of $X_1,...,X_4$, $c$ is summed over all possible fusion channels of $X_1$ and $X_2$ and $\bar{c}$ is the conjugate of $c$. Thus, when the Josephson junction pins the fusion channel $a$ in the $X_2\times X_3$ basis, the interferometer sees a superposition of $X_1\times X_2$ fusion channels.

The loop operator associated with a tunneling anyon $b$ acts diagonally in the $X_1\times X_2$ basis,
\begin{align}
   W_{N_{\text{qh},b}}\ket{X_1\times X_2=c,~X_3\times X_4=\bar{c}}=M_{bc}\ket{X_1\times X_2=c,~X_3\times X_4=\bar{c}}\,,
\end{align}
where $M_{bc}$ is the monodromy between anyon $b$ and $c$. Therefore, if the Josephson junction pins the $X_2\times X_3$ fusion channel to $a$, the expectation value of the loop operator appearing in Eq.~\eqref{eqn:general_fp} is
\begin{align}
    \ev{W_{N_{\text{qh}},b}}_a=\sum_{c}\abs{\pbr{F_{XXX}^X}_{ac}}^2M_{bc}\,.
\end{align}
Thus, the interference term is a weighted sum over the $X_1\times X_2$ fusion channels, with weights determined by the corresponding $F$-symbols. 

For the fermionic Laughlin FQH-SC at filling $\nu=1/m$, the defect fusion channels are labeled by $k\in\Z_{2m}$. The relevant $F$-symbol \cite{tambara-1998} and monodromy are
\begin{align}
    \sbr{F_{XXX}^{X}}_{kr}=\frac{1}{\sqrt{2m}} e^{-\frac{2\pi i}{2m}kr}\,,\quad M_{jr}=e^{\frac{2\pi i}{m}jr}\,.
\end{align}
Here $k,r\in\Z_{2m}$ are defect fusion-channel labels and $j\in\Z_m$ is the label of the tunneling anyon in the Laughlin FQH state. The expectation value of the loop operator is
\begin{align}
    \ev{W_{N_{\text{qh},j}}}_r=\frac{1}{2m}\sum_{k=0}^{2m-1}e^{\frac{2\pi i}{m}jk}=\de_{j,0\text{ mod } m}\,.
\end{align}
This means that for tunneling of any nontrivial anyon in the interferometer, the interference vanishes, giving the longitudinal conductance
\begin{align}
    \si_{xx}\propto \abs{\Ga_1}^2+\abs{\Ga_2}^2\,.
\end{align}
The same cancellation occurs for Abelian FQH-SCs in general. The $F$-move gives an equal-weight sum over the Abelian defect fusion channels, and the monodromy phases of a nontrivial tunneling quasiparticle average to zero. This conclusion is independent of whether fermion parity contributes to the zero mode Hilbert-space dimension.

For the fermionic MR FQH-SC at filling $\nu=1/m$, the neutral and charge sectors factorize. The relevant $F$-symbol \cite{tambara-1998} is
\begin{align}
    \sbr{F_{XXX}^{X}}_{(k_n,k)(r_n,r)}=\frac{1}{\sqrt{4m}} e^{-\frac{2\pi i}{2}k_nr_n}e^{-\frac{2\pi i}{2m}kr}\,.
\end{align}
Here $k_n,r_n\in\Z_2^{(n)}$ and $k,r\in\Z_{2m}$ are defect fusion-channel labels. The same conclusion holds for the MR FQH-SC. Because the defect $F$-move factorizes into neutral and charge sectors with equal weights, the weighted sum over $(k_n,k)$ averages the corresponding monodromy phases to zero for a nontrivial tunneling quasiparticle. Thus, after summing over the fusion channels selected by the basis transformation, the interferometric cross term cancels in the same way as in the Abelian cases. 
\end{document}